\documentclass[%
twocolumn,
groupedaddress,
amsmath,amssymb,
aps,
pre,
]{revtex4-2}

\usepackage{graphicx}
\usepackage{amsfonts}
\usepackage{bm}
\usepackage{hyperref,url}
\usepackage{changes}
\usepackage[dvipsnames]{xcolor}
\usepackage{soul}
\usepackage{dcolumn}

\begin{document}


\title{Wavelet-based multiresolution analysis of quantum fractals in confined dynamics}

\author{David Navia}
\affiliation{Department of Optics, Faculty of Physical Sciences, Universidad
Complutense de Madrid\\
Pza.\ Ciencias 1, Ciudad Universitaria -- 28040 Madrid, Spain}

\author{\'Angel S. Sanz}
\affiliation{Department of Optics, Faculty of Physical Sciences, Universidad
Complutense de Madrid\\
Pza.\ Ciencias 1, Ciudad Universitaria -- 28040 Madrid, Spain}

\date{\today}

\begin{abstract}
Fractal structures naturally emerge in quantum systems whose initial states exhibit spatial discontinuities, a phenomenon first identified by Berry in the paradigmatic case of a particle confined in an infinite potential well.
While previous analyses of quantum fractals have mainly relied on spectral decompositions and geometric scaling arguments, their quantitative characterization often depends on scale choices and truncation effects.
Here we present a wavelet-based multiresolution framework that enables a direct and assumption-free quantification of quantum fractality.
Fractal dimensions are extracted from the scale-dependent distribution of wavelet energies, without invoking prior power-law hypotheses.
The method is applied to space and time quantum fractals arising in confined dynamics, as well as to dynamical curves generated by the associated quantum probability flux.
These flux-driven trajectories provide a natural space--time parametrization of the underlying fractal structure and yield scaling properties fully consistent with Berry's predictions for space--time fractals.
The resulting fractal dimensions are shown to be robust with respect to the choice of wavelet family, numerical cutoffs, and system parameters.
Beyond validating earlier conjectures, the present framework offers a unified and computationally efficient tool for the multiscale analysis of quantum fractality in confined and interference-driven quantum dynamics.
That is, it provides an operational, scale-adaptive criterion that unifies the characterization of space, time, and space--time quantum fractals within a single, hypothesis-free approach.
\end{abstract}


\maketitle


\section{Introduction}
\label{sec1}

Fractal geometry provides a powerful framework for describing scale-invariant features across a wide range of physical systems, from classical chaotic dynamics to transport phenomena and wave propagation in complex media \cite{mandelbrot-bk:1983,feder-bk}.
In quantum mechanics, the emergence of fractal structures is particularly striking due to
the constraints imposed by the uncertainty principle and the expected regularity of quantum
states.
Nevertheless, Berry showed that quantum states evolving from initial conditions with spatial
discontinuities can naturally develop self-similar structures in space, time, and joint
space--time domains \cite{berry:JPA:1996}.
Recently, quantum fractality has been revisited from a broader perspective across different platforms
and contexts \cite{bercioux:naturePhysics:2019}, including experimental realizations of electronic states
in fractal geometries \cite{kempkes:naturePhysics:2019} and wave dynamical studies of quantum transport in fractal networks \cite{xu:NatPhotonics:2021}.

The paradigmatic example of a square wave function confined in an infinite potential well
reveals the formation of quantum carpets whose spatial and temporal profiles exhibit fractal
features at irrational fractions of the recurrence time.
Berry further conjectured that the corresponding fractal dimensions take universal values,
namely $D_{\mathrm{space}} = 3/2$ and $D_{\mathrm{time}} = 7/4$, independent of the system
parameters governing the evolution.
These predictions have motivated extensive numerical and theoretical investigations of
quantum fractals in confined and unconfined systems
\cite{bialynicki:PhysicaD:1994,wojcik:PRL:2000,robinett:PhysRep:2004}.

Quantitative characterizations of quantum fractality have traditionally relied on geometric
constructions—such as box-counting or arc-length scaling—or on spectral power-law analyses
based on Fourier decompositions \cite{mandelbrot-bk:1983,feder-bk,BerryLewis:PRSA:1980,falconer-bk:1990,wornell-bk:1996}.
While these approaches successfully capture the asymptotic scaling behavior, they often rely
on implicit assumptions concerning scaling ranges, are sensitive to finite-size and
truncation effects, and typically require high-resolution sampling.
The emphasis here is not on obtaining new asymptotic values, but on introducing a diagnostic that remains reliable in the pre-fractal regime imposed by spectral truncation, where standard geometric or Fourier-based methods become ambiguous.

Unlike approaches based on multifractal spectrum reconstruction or power-law fits within predefined spectral representations \cite{wornell-bk:1996,kantelhardt:PhysicaA:2002}, the approach adopted here is based on wavelet multiresolution analysis.
Wavelets provide a localized and scale-adaptive decomposition that allows scaling properties
to be extracted directly from the distribution of energy across dyadic scales, without prior
assumptions about the form, extent, or location of the scaling laws.
This makes them particularly well suited for the analysis of quantum fractals, which are
intrinsically affected by numerical cutoffs and pre-fractal behavior.

Specifically, in this work we have applied wavelet-based multiresolution analysis to the study of quantum fractals arising in confined dynamics.
We analyze fractal structures manifested in spatial and temporal profiles of the probability
density and extend the analysis to dynamical curves generated by the associated quantum
probability flux.
These \mbox{flux-driven} trajectories, determined within the Bohmian picture of \mbox{quantum mechanics} \cite{sanz:AJP:2012,sanz:FrontPhys:2019}, provide a natural and non-arbitrary parametrization of the space--time organization of the quantum state, offering an alternative to diagonal cuts through quantum carpets while remaining fully consistent with Berry's predictions for space--time fractals.
Thus, although the contribution of this work is primarily methodological, it directly addresses a long-standing practical limitation in the quantitative analysis of quantum fractals.

Accordingly, this work has been organized as follows.
Section~\ref{sec2} introduces the essential elements of wavelet multiresolution analysis and summarizes the confined quantum model, including flux-based trajectories as space--time probes.
The main numerical results are presented and discussed in Sec.~\ref{sec3}, and concluding remarks are given in Sec.~\ref{sec4}.


\section{Theoretical framework}
\label{sec2}


\subsection{Wavelet multiresolution analysis and fractal scaling}
\label{sec21}

Wavelet multiresolution analysis provides a natural framework for characterizing scale-invariant structures in signals that exhibit nonstationary or localized features \cite{mallat:IEEE:1989,daubechies:CommunPureApplMath:1988}.
Unlike Fourier-based approaches, which rely on global basis functions, wavelets offer a
simultaneous localization in both scale and position, making them particularly suited for the
analysis of fractal and pre-fractal structures \cite{Gabor:JIEE:1946,wornell-bk:1996,mallat:Elsevier:2009}.

Within the discrete wavelet transform, a signal $f(n)$ sampled over $2^M$ points can be
decomposed into approximation and detail coefficients across a hierarchy of dyadic scales (see Appendix~\ref{appA} for further details).
At each scale $j$, the wavelet detail coefficients $d_{j,k}$ quantify the fluctuations of the signal at that resolution.
The energy associated with scale $j$ is defined as
\begin{equation}
E_j = \sum_k |d_{j,k}|^2 ,
\end{equation}
from which a scale-dependent energy distribution can be constructed.

For statistically self-similar signals, the wavelet energies follow a power-law scaling,
\begin{equation}
E_j \sim 2^{-j\alpha},
\end{equation}
where $\alpha$ is related to the Hurst exponent $H$ [see Eq.~\eqref{vardjk}].
In practice, the scaling exponent is obtained from a linear regression of $\log_2 E_j$ versus $j$ over an intermediate range of scales, excluding both the coarsest levels (finite-size effects) and the finest levels (truncation/noise).
Uncertainties reported in the Supplemental Material (\mbox{Tables}~\mbox{S1--S9}) correspond to the standard error of the fitted slope within that scaling window.

For an orthonormal discrete wavelet transform, $E_j=\sum_k |d_{j,k}|^2$ is proportional to the
variance of the detail coefficients at level $j$; therefore, the fitted slope in $\log_2 E_j$
versus $j$ directly yields $H$ via Eq.~\eqref{vardjk}.
The fractal dimension of the corresponding curve is then obtained from
\begin{equation}
D = 2 - H .
\label{fractdim}
\end{equation}
Related discussions on parameter choices and finite-sample effects in wavelet-based Hurst estimation can be found in \cite{Wu:entropy:2020}; see also \cite{KnightNasonNunes:statcomput:2017} for wavelet-based long-memory estimation in modern settings.

A key advantage of this approach is that the scaling exponent is extracted directly from the
distribution of energy across scales, without requiring prior assumptions about scaling ranges
or geometric constructions.
This makes wavelet analysis particularly robust in the presence of finite-size effects and
numerical cutoffs, which are intrinsic to quantum pre-fractals.
Further details on the selection of the scaling window and on uncertainty estimation are provided in the Supplemental Material.

\begin{figure*}[!t]
 \centering
 \includegraphics[width=\textwidth]{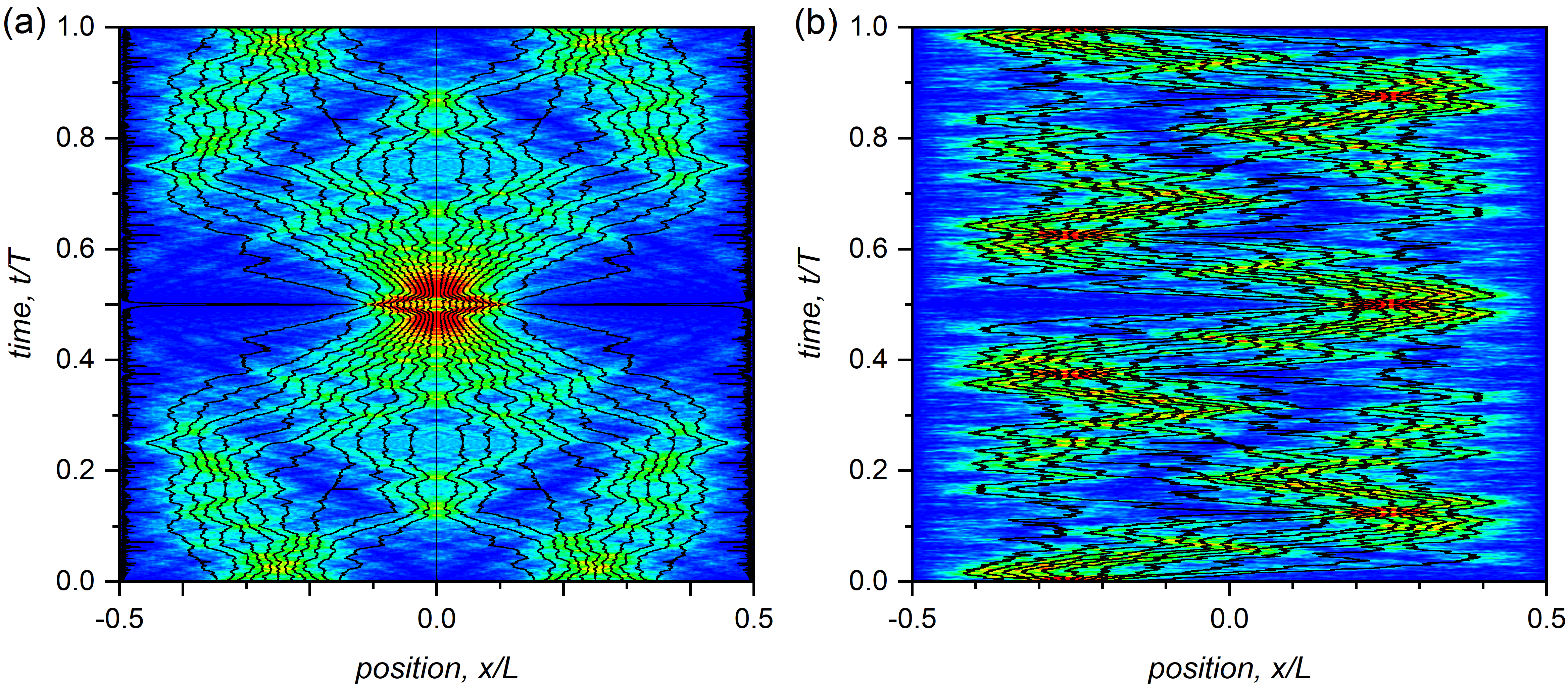}
 \caption{\label{Fig1}
  Fractal space--time quantum carpets generated by the evolution in time of the probability density associated with (a) a coherent superposition of two symmetrically distributed square functions, and (b) only the left-hand side square function.
  In both cases, the width of the square functions is such that $w/L = 0.25$ and are centered at $x_\pm = \pm 0.25 L$.
  In the density plots, blue denotes lower values of the probability density, while the transition to red indicates increasing values (for a better visualization of the details both probability densities )
  For a better visualization of the carpets' details, the probability densities have been truncated to about $60\%$ their maxima.
  To make evident the underlying local fractal dynamics, sets of flux-based (Bohmian) trajectories are superimposed to the quantum carpets (black solid lines).
  Their initial positions are taken evenly spaced in those regions with non-vanishing probability density.}
\end{figure*}


\subsection{Quantum fractals in confined dynamics}
\label{sec22}

We consider a quantum particle confined within a one-dimensional infinite potential well (see Appendix~\ref{appB} for further details).
When the initial state exhibits spatial discontinuities, such as in the case of square wave
packets, the subsequent quantum evolution generates intricate interference patterns known
as quantum carpets.
At irrational fractions of the recurrence time, spatial profiles of the probability density
display fractal characteristics, while temporal profiles at fixed positions also exhibit fractal behavior.

Although the full quantum evolution is well defined in terms of spectral decompositions \cite{sanz:PhysScr:2019}, the associated fractal structures correspond to pre-fractal realizations limited by the finite number of contributing eigenstates.
This feature makes the system an ideal testbed for multiresolution methods capable of
capturing scaling behavior in the presence of natural cutoffs.

In the present work, space and time quantum fractals are treated on equal footing within
the wavelet framework, allowing their fractal dimensions to be quantified in a unified manner.
The analysis does not rely on geometric constructions or assumed power laws, but instead
extracts scaling information from the multiscale organization of the probability density.


\subsection{Flux-based trajectories as space--time probes}
\label{sec23}

In addition to spatial and temporal profiles of the probability density, quantum fractality can also be explored through dynamical curves generated by the quantum probability flux associated with the evolving state.
Here, fluxed-based trajectories in the framework of the Bohmian picture of quantum mechanics \cite{sanz:FrontPhys:2019} are considered as operational tools that provide a continuous space--time parametrization of the probability flow and serve as probes of the underlying fractal structure \cite{sanz:JPA:2005}.
Rather than probing quantum carpets along a priori chosen geometric cuts, which implicitly privilege specific space--time directions, we employ flux-based trajectories as adaptive probes that dynamically select the regions where space and time correlations actually develop.
This provides a non-arbitrary way to interrogate space--time correlations in a manner consistent with Berry's predictions for space--time fractals \cite{berry:JPA:1996}.

The trajectories are obtained by integrating the local velocity field defined as the ratio
between the probability current density $j(x,t)$ and the probability density $\rho(x,t)$ (see Appendices~\ref{appB} and \ref{appC} for further details).
This construction is formally equivalent to the trajectory formulation commonly referred to as
the Bohmian formulation of quantum mechanics.
In the present context, the term \emph{Bohmian trajectories} is used solely as a conventional
label for these flux-based curves, without invoking any interpretative assumptions regarding
their ontological \mbox{status.}

By following the evolution of the probability flow, these trajectories encode space--time
correlations of the quantum dynamics in a non-arbitrary manner.
As such, they provide a dynamically motivated alternative to geometric diagonal cuts through
quantum carpets and yield scaling properties consistent with those predicted by Berry for
space--time quantum fractals.
Their role in this work is therefore complementary: they supply an additional, flow-based
perspective that reinforces the multiresolution analysis of quantum fractality.
It is worth stressing that the specific form of the trajectories is not essential here; their role is purely operational, serving as flow-following parametrizations of the quantum dynamics rather than as physical entities.


\section{Results}
\label{sec3}


\subsection{Space quantum fractals}

We begin by applying the wavelet multiresolution analysis to space quantum fractals, which
emerge in the probability density at irrational fractions of the recurrence time.
Figure~\ref{Fig1} shows the quantum carpets generated by two wave functions that give rise to the appearance of quantum fractals.
To the left, Fig.~\ref{Fig1}(a) shows the symmetric carpet associated with an initial wave function consisting of the coherent superposition of two square amplitudes, each allocated on one half of the box (see details in the figure caption).
An asymmetric configuration is shown to the right, in Fig.~\ref{Fig1}(b), where the initial wave function only consists of the left-hand side square.
Following the detailed analysis in \cite{sanz:PhysScr:2019}, the revival time in each case is going to be different, being larger in the second case.

\begin{figure}[t]
 \centering
 \includegraphics[width=\columnwidth]{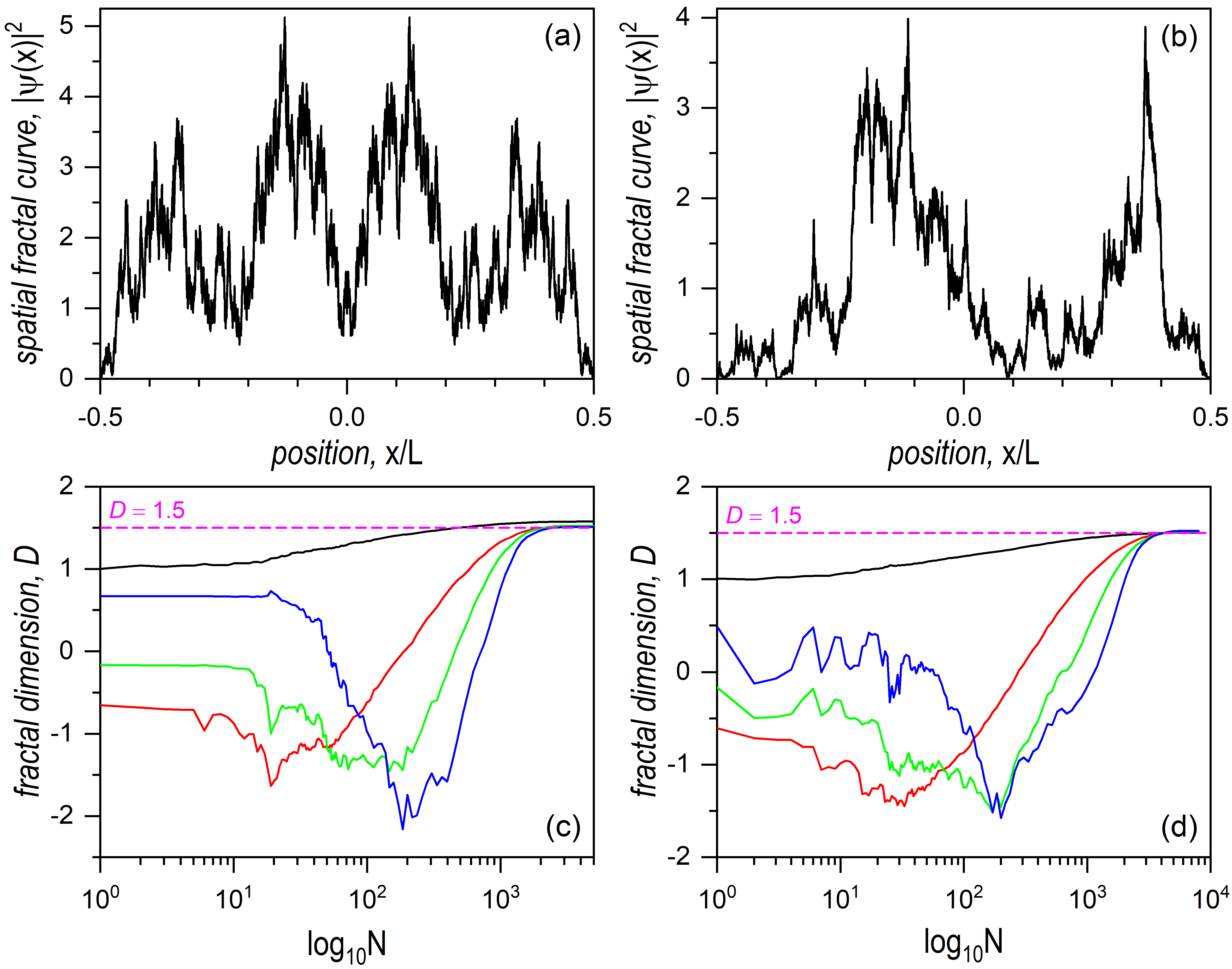}
 \caption{\label{Fig2}
  Space quantum fractal profile exhibited by the probability density of the symmetric (a) and asymmetric (b) wave functions shown in Fig.~\ref{Fig1}.
  Both space fractals are computed at $t = T/\sqrt{2}$, where $T$ denotes the revival time corresponding to each case.
  The numerical estimates of the corresponding fractal dimensions obtained from the multiresolution analysis are shown, respectively, in panels (c) and (d): Haar (black), db4 (red), db8 (green), and db16 (blue).
  These estimates are determined from convergence with the number of basis functions considered in the spectral reconstruction without adding any extra hypothesis.
  As it is shown, beyond $N=1,\!000$, all results show fast convergence to the value $D = 1.5$ (magenta dashed line) conjectured by Berry for space fractals \cite{berry:JPA:1996}.}
\end{figure}

Figure~\ref{Fig2} shows representative spatial profiles for both symmetric (a) and asymmetric (b) initial states, evaluated at $t = T/\sqrt{2}$, where $T$ is revival time \cite{sanz:PhysScr:2019} corresponding to each case: $T = (2\pi)^{-1} \approx 0.16$ for the symmetric quantum carpet, and $T = 4/\pi \approx 1.27$ for the asymmetric one.
In both cases, the probability density exhibits the characteristic irregular structure associated with quantum fractality.

The corresponding fractal dimension estimates, obtained from the scale-dependent distribution
of wavelet energies, are displayed in Fig.~\ref{Fig2}, in panels (c) and (d), respectively.
For increasing spectral truncation $N$, all wavelet families considered converge towards the
value $D \simeq 3/2$, in agreement with Berry's theoretical prediction for space quantum
fractals.
While short-support wavelets (db1) tend to display smoother convergence, higher-order Daubechies wavelets exhibit faster stabilization once sufficient scale separation is achieved.

These results demonstrate that the wavelet-based method yields robust and wavelet-independent
fractal dimensions, even in the presence of finite-size effects and pre-fractal behavior inherent to truncated quantum evolutions.

\subsection{Time quantum fractals}

\begin{figure}[t]
 \centering
 \includegraphics[width=\columnwidth]{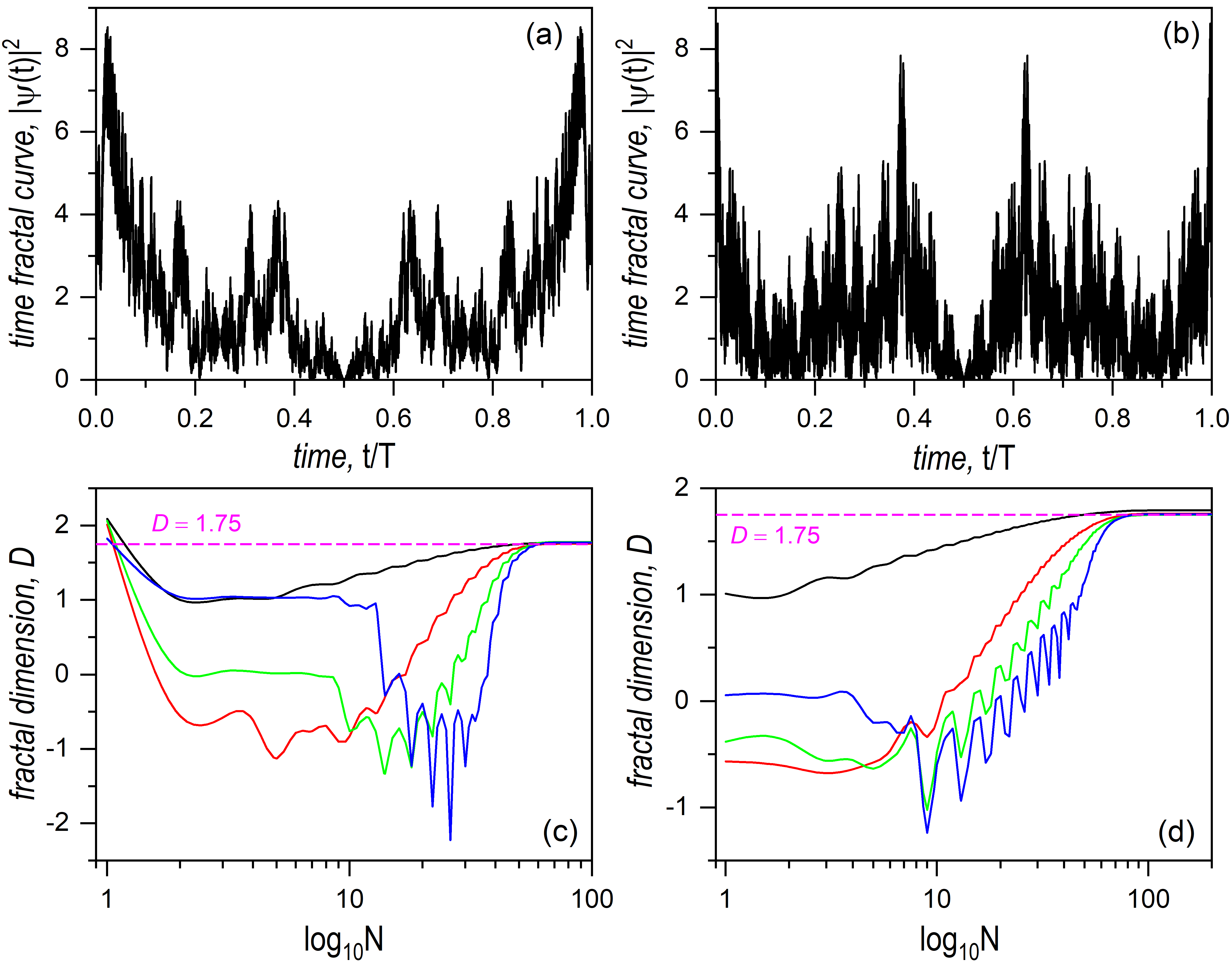}
 \caption{\label{Fig6}
  Time quantum fractal profile exhibited by the probability density of the symmetric (a) and asymmetric (b) wave functions shown in Fig.~\ref{Fig1}.
  Both time fractals are computed at $x = -0.25 L$, where $L$ denotes the total length of the cavity.
  The numerical estimates of the corresponding fractal dimensions obtained from the multiresolution analysis are shown, respectively, in panels (c) and (d): Haar (black), db4 (red), db8 (green), and db16 (blue).
  These estimates are determined from convergence with the number of basis functions considered in the spectral reconstruction without adding any extra hypothesis.
  As it is shown, beyond $N \gtrsim 100$, all results show fast convergence to the value $D = 1.75$ (magenta dashed line) conjectured by Berry for time fractals \cite{berry:JPA:1996}.}
\end{figure}

We next consider time quantum fractals, obtained by following the probability density at a
fixed space position over one recurrence period.
Figure~\ref{Fig6} shows representative temporal profiles for symmetric (a) and asymmetric (b) initial conditions.
Except for nodal points dictated by symmetry, all temporal traces display fractal fluctuations
across multiple time scales.

The corresponding fractal dimension estimates are summarized in
Fig.~\ref{Fig6}, in panels (c) and (d), respectively.
In contrast to the spatial case, convergence towards the asymptotic value $D \simeq 7/4$ is
typically achieved for smaller values of $N$, reflecting the fact that time fractality is
present at all positions.
As before, the estimates are largely insensitive to the choice of wavelet basis, highlighting
the robustness of the multiresolution framework.

\subsection{Flux-based trajectories and space--time fractality}

Finally, we extend the analysis to flux-based trajectories generated by the quantum probability
current.
Figure~\ref{Fig1} also shows representative trajectories superimposed on the
corresponding quantum carpets for symmetric and asymmetric initial states.
These curves provide a continuous space--time parametrization of the evolving probability
flow and naturally explore the regions where space and time fractal features intertwine.

\begin{figure}[!t]
  \centering
  \includegraphics[width=\columnwidth]{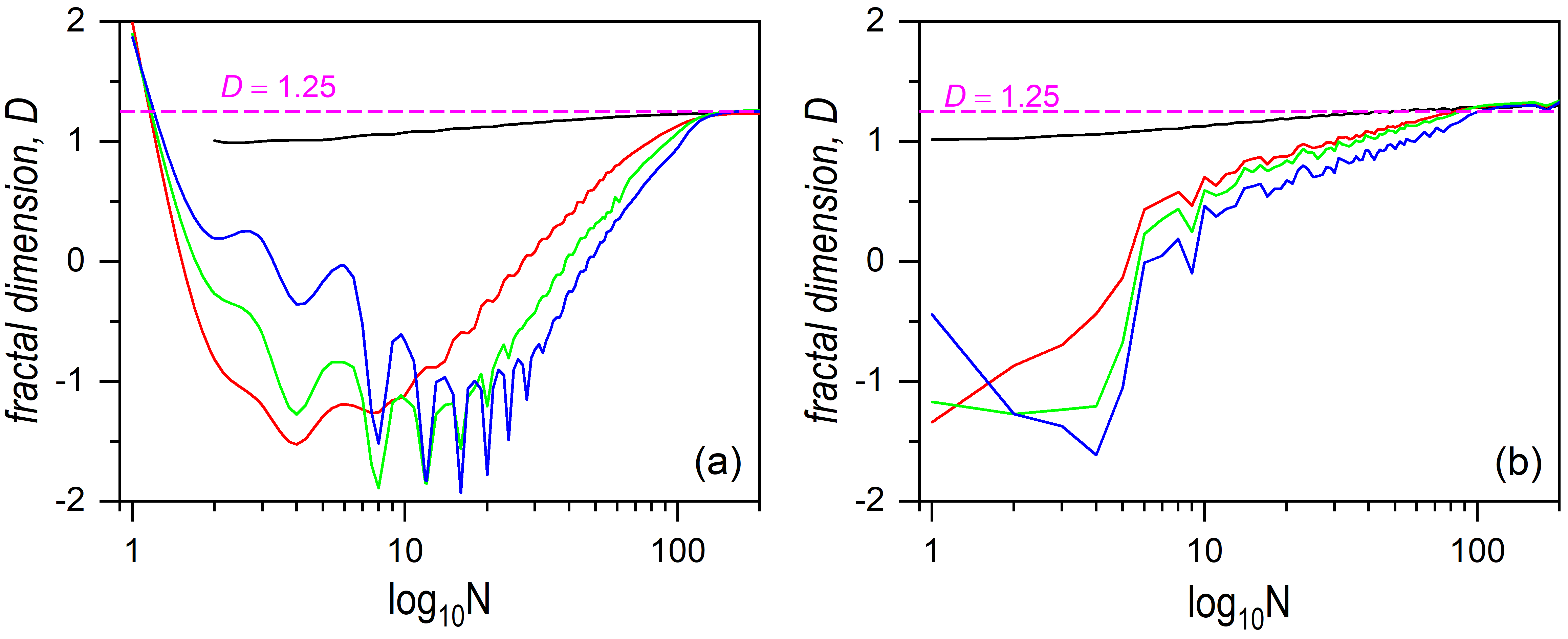}
  \caption{\label{Fig8}
   Numerical estimates of the fractal dimension obtained from the multiresolution analysis for three flux-based fractal (Bohmian) trajectories for different instances: Haar (black), db4 (red), db8 (green), and db16 (blue).
   The results in panels (a) and (b) correspond to trajectories with initial conditions taken at $x_0 = -0.25 L$ associated with the symmetric and asymmetric wave functions whose carpets are shown in Figs.~\ref{Fig1}(a) and (b), respectively, with $w/L = 0.25$.
   Note that, beyond $N \gtrsim 100$, all results show fast convergence to the value $D = 1.25$ (magenta dashed line) conjectured by Berry for space--time fractals \cite{berry:JPA:1996}.}
\end{figure}

The wavelet analysis of individual trajectories reveals clear scaling behavior over a broad
range of resolutions.
Figure~\ref{Fig8} displays the convergence of the estimated fractal dimension as a function of spectral truncation.
In all cases considered, the asymptotic value approaches $D \simeq 5/4$, consistent with Berry's prediction for space--time quantum fractals.

Unlike geometric diagonal cuts through quantum carpets, which must be chosen a priori and may miss dynamically relevant structures, flux-based trajectories automatically adapt to the evolving interference pattern, as they encode space--time correlations in a dynamically motivated manner, without relying on arbitrary directions.
Their agreement with the known fractal dimensions reinforces the interpretation of these
curves as effective probes of quantum space--time fractality, complementary to standard
space and time analyses.
Indeed, this agreement indicates that the trajectories capture the same underlying space--time scaling properties as full carpet analyses, but with a reduced and dynamically motivated sampling.

\section{Conclusions}
\label{sec4}

Quantum states evolving from discontinuous initial conditions provide a natural setting for
the emergence of fractal structures in confined dynamics.
Since Berry's seminal work, the existence of space, time, and space--time quantum
fractals has been firmly established, together with the conjecture that their associated
fractal dimensions take universal values.
Despite this progress, their quantitative characterization has remained closely tied to
spectral decompositions and geometric scaling arguments, which are often sensitive to
finite-size effects and numerical cutoffs.
The present work shows that these structures can be quantitatively characterized in a unified and numerically stable manner, even under the unavoidable limitations imposed by finite spectral resolution.

In this work, we have introduced a wavelet-based multiresolution framework for the analysis
of quantum fractality that circumvents these limitations.
By extracting scaling properties directly from the scale-dependent distribution of wavelet
energies, fractal dimensions can be determined without invoking prior power-law assumptions
or arbitrary geometric constructions.
Applied to space and time quantum fractals in confined systems, the method yields
robust and wavelet-independent estimates fully consistent with Berry's theoretical
predictions.

Beyond standard space and time analyses, we have shown that dynamical curves generated
by the quantum probability flux provide a natural space--time parametrization of quantum
fractals.
These flux-based trajectories act as effective probes of space--time correlations in the
quantum dynamics and reproduce the fractal dimensions associated with space--time quantum
fractals, without recourse to diagonal cuts through quantum carpets.
Their role in the present work is methodological: they complement the multiresolution analysis
by supplying a dynamically motivated perspective that reinforces, rather than replaces,
the established description of quantum fractality.

The framework presented here is general and readily applicable to other quantum systems in
which fractal or multiscale features arise, including interference-driven dynamics,
wave propagation in confined geometries, and optical analogues in paraxial regimes.
More broadly, the combination of multiresolution analysis with flow-based probes provides a practical framework for diagnosing quantum fractality in systems where analytical access is limited and numerical implementations are necessarily pre-fractal.
We thus expect that these tools will prove useful in the study of more intricate quantum systems
exhibiting anomalous transport, critical dynamics, or emergent complexity, where a unified
and scale-adaptive perspective is essential.



\section*{Acknowledgments}

This research has been funded by the ``European Union NextGenerationEU/PRTR and MCIN/AEI/10.13039/
501100011033'' Project reference \mbox{PID2021-127781NB-I00}.


\section*{Data availability}

The data that support the findings of this article are not publicly available.
The data are available from the authors upon reasonable request.


\appendix


\section{Wavelet multiresolution analysis and fractal scaling}
\label{appA}

In this appendix we summarize the technical aspects of the wavelet multiresolution analysis
used throughout the main text, following standard formulations \cite{mallat:IEEE:1989,daubechies:CommunPureApplMath:1988}.

A discrete signal $f(n)$ sampled over $2^M$ points can be decomposed into approximation and
detail coefficients via an orthonormal wavelet basis,
\begin{equation}
f(n) = \sum_k a_{J,k}\,\phi_{J,k}(n) + \sum_{j=1}^{J}\sum_k d_{j,k}\,\psi_{j,k}(n),
\end{equation}
where $\phi_{J,k}$ and $\psi_{j,k}$ denote scaling and wavelet functions, respectively.

The energy associated with scale $j$ is defined as
\begin{equation}
E_j = \sum_k |d_{j,k}|^2,
\end{equation}
from which the total energy follows as $E_{\mathrm{tot}}=\sum_j E_j$.

For statistically self-similar signals, the wavelet coefficient variance exhibits power-law
scaling with scale,
\begin{equation}
\mathrm{Var}(d_{j,k}) \sim 2^{j(2H+1)},
\label{vardjk}
\end{equation}
where $H$ denotes the Hurst exponent.
The corresponding fractal dimension of the graph is then obtained via the relation $D = 2 - H$, that is, Eq.~\eqref{fractdim}.


\section{Spectral representation and probability flux}
\label{appB}

The quantum dynamics considered in this work is based on the spectral decomposition of the
wave function in the eigenbasis of the infinite potential well,
\begin{equation}
\Psi(x,t) = \sum_{n=1}^{\infty} c_n \phi_n(x) e^{-iE_n t/\hbar},
\end{equation}
where $\phi_n(x)$ are the energy eigenfunctions and $c_n$ the expansion coefficients
determined by the initial condition.

For discontinuous initial states, such as square wave packets, the coefficients decay
algebraically as $|c_n|\sim n^{-2}$, leading to divergent energy expectation values and the
emergence of pre-fractal behavior.

The probability density and current density are defined in the usual manner as
\begin{align}
\rho(x,t) &= |\Psi(x,t)|^2, \\
j(x,t) &= \frac{\hbar}{m}\,\mathrm{Im}\!\left[\Psi^*(x,t)\,\partial_x\Psi(x,t)\right].
\end{align}
These quantities satisfy the continuity equation,
\begin{equation}
\partial_t \rho(x,t) + \partial_x j(x,t) = 0.
\end{equation}


\section{Flux-based trajectories and standard constructions}
\label{appC}

The flux-based trajectories considered in this work follow from the local velocity field \cite{sanz:FrontPhys:2019}, defined as
\begin{equation}
v(x,t) = \frac{j(x,t)}{\rho(x,t)}.
\end{equation}
Integrating this field yields continuous curves $x(t)$ that follow the probability flow
associated with the evolving quantum state.
From a numerical standpoint, this construction also reduces the effective dimensionality of the sampling problem, providing a computationally efficient access to space--time scaling properties.

This construction is equivalent to the trajectory formulation commonly referred to as the
Bohmian formulation of quantum mechanics \cite{sanz:AJP:2012,sanz:FrontPhys:2019}.
In the present work, this equivalence is used exclusively as a technical reference, without
attaching any interpretative significance to the individual trajectories.

The purpose of including these flux-based curves is to provide a dynamically motivated
space--time parametrization of quantum fractals, thereby complementing purely geometric
analyses of quantum carpets.




%



\renewcommand{\thefigure}{S\arabic{figure}}
\renewcommand{\thetable}{S\arabic{table}}
\setcounter{table}{0}
\setcounter{figure}{0}

\begin{widetext}

\section*{SUPPLEMENTAL MATERIAL}
This Supplemental Material reports additional numerical results, including fitted scaling exponents and their standard errors (reported as $\pm$ values), for multiple wavelet families, system parameters, and initial conditions.

\subsection*{Fractal geometry and wavelet-based multiresolution analysis}

\begin{figure*}[!b]
  \centering
  \includegraphics[width=0.85\textwidth]{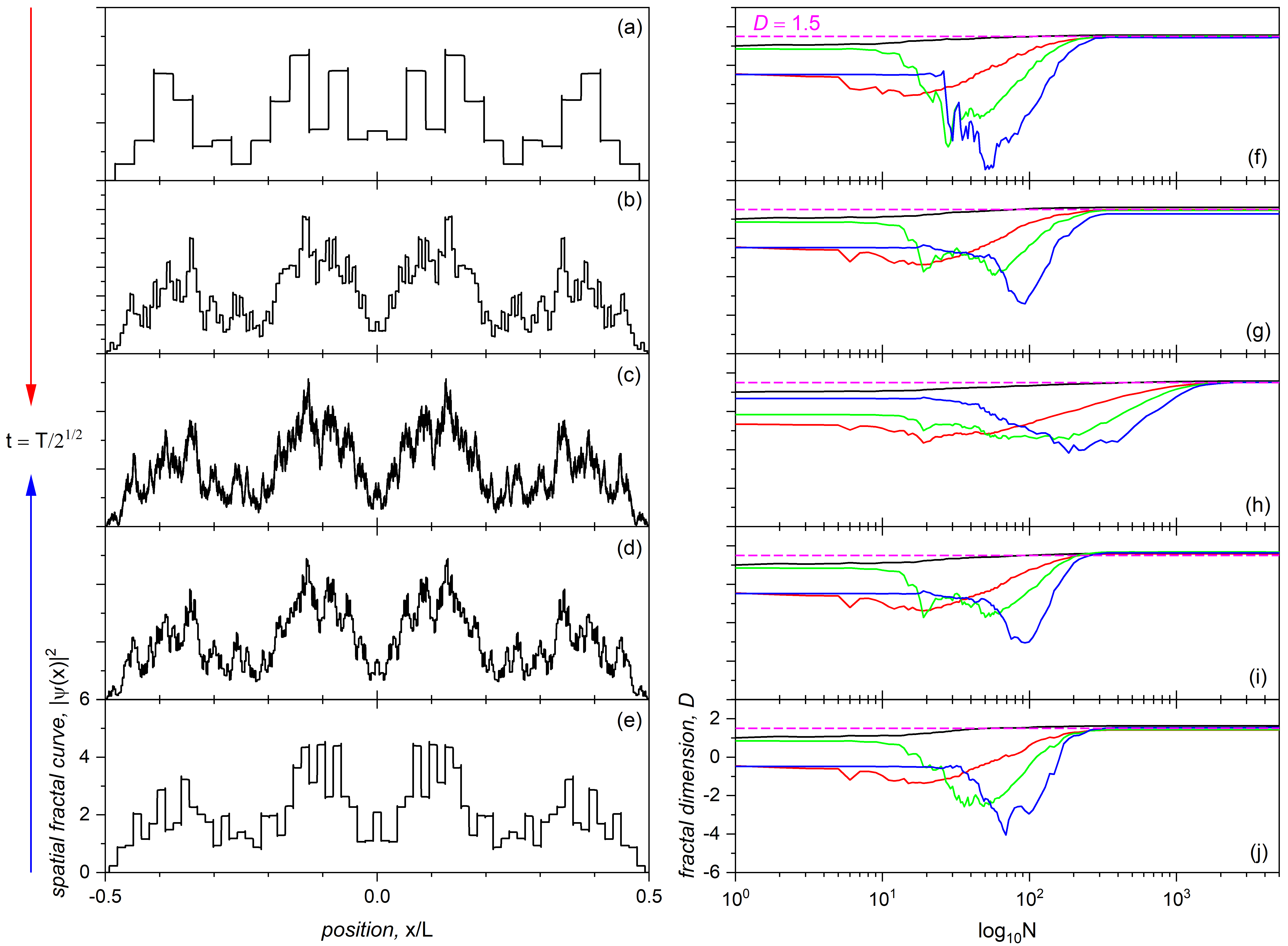}
  \caption{\label{FigS1} Panels (a) to (e) show the convergence to the space quantum fractal observable at $t = T/\sqrt{2}$ [see Fig.~2(a) in main text] from higher and lower commensurate fractions of the recurrence time: (a) $t = 5 T/7 \approx 0.714286 T$, (b) $t = 29 T/41 \approx 0.707317 T$, (c) $t = T/\sqrt{2} \approx 0.707107 T$ [same as in Fig.~2(a)], (d) $t = 70T /99 \approx 0.707071 T$, and (e) $t = 12 T/17 \approx 0.705882 T$, with $T = (2\pi)^{-1}$.
  The numerical estimates of the corresponding fractal dimension rendered by the wavelet-based multiresolution analysis here considered are shown in panels (f) to (j): Haar (black), db4 (red), db8 (green), and db16 (blue).
  Although only curve in panel (c) is a fractal, note that the multiresolution analysis provides us with the same number in all other cases, which should not be interpreted here in a strict sense, as a measure of the fractal geometry, but also, in a broader sense, as a measure of the capacity that a series expansion has to eventually generate a fractal.
  In this case, the fractal dimension $D = 1.5$ [magenta dashed line in panels (f) to (j)] corresponds to this capacity.
  In this sense, any series with $D = 1$ will generate no fractal structure at any time.}
\end{figure*}

\clearpage

\subsection*{Wavelet-based estimates of the fractal dimension $D$ in various instances}

For each entry in the tables below, $D$ has been obtained from a linear regression of $\log_2 E_j$ versus $j$ within the scaling window defined in Sec.~II.A (in the main text), excluding the coarsest scales dominated by finite-size effects and the finest scales dominated by truncation/noise. Quoted uncertainties ($\pm$) correspond to the standard error of the fitted slope propagated to $D$.

\begin{table}[h]
\centering
\caption{\label{tab1} Fractal dimension for space quantum fractals profiles extracted from Fig.~1(a) at the times $t$ indicated in the first column.}
\begin{tabular}{ccccc}
Time & haar & db4 & db8 & db16 \\
\hline
$T/\sqrt{2}$ & 1.559 $\pm$ 0.000 & 1.447 $\pm$ 0.023 & 1.455 $\pm$ 0.026 & 1.433 $\pm$ 0.000 \\
$T/\sqrt{3}$ & 1.407 $\pm$ 0.010 & 1.468 $\pm$ 0.010 & 1.399 $\pm$ 0.000 & 1.443 $\pm$ 0.000 \\
$T/\sqrt{5}$ & 1.504 $\pm$ 0.005 & 1.531 $\pm$ 0.006 & 1.558 $\pm$ 0.000 & 1.584 $\pm$ 0.000 \\
$T/\sqrt{7}$ & 1.386 $\pm$ 0.025 & 1.480 $\pm$ 0.010 & 1.506 $\pm$ 0.034 & 1.468 $\pm$ 0.025 \\
$T/\sqrt{10}$ & 1.433 $\pm$ 0.012 & 1.417 $\pm$ 0.022 & 1.484 $\pm$ 0.008 & 1.482 $\pm$ 0.026 \\
\end{tabular}
\end{table}

\begin{table}[h]
\centering
\caption{\label{tab5} Fractal dimension for time quantum fractals profiles extracted from Fig.~1(a) at the fixed $x-$positions indicated in the first column.}
\begin{tabular}{ccccc}
$x-$position & haar & db4 & db8 & db16 \\
\hline
$-0.4167$ & 1.751 $\pm$ 0.002 & 1.735 $\pm$ 0.001 & 1.731 $\pm$ 0.006 & 1.721 $\pm$ 0.004 \\
$-0.3333$ & 1.761 $\pm$ 0.006 & 1.754 $\pm$ 0.011 & 1.733 $\pm$ 0.004 & 1.735 $\pm$ 0.003 \\
$-0.2500$ & 1.757 $\pm$ 0.002 & 1.753 $\pm$ 0.008 & 1.754 $\pm$ 0.006 & 1.792 $\pm$ 0.000 \\
$-0.1667$ & 1.752 $\pm$ 0.006 & 1.744 $\pm$ 0.003 & 1.744 $\pm$ 0.002 & 1.778 $\pm$ 0.002 \\
$-0.0833$ & 1.766 $\pm$ 0.001 & 1.773 $\pm$ 0.004 & 1.805 $\pm$ 0.003 & 1.738 $\pm$ 0.045 \\
$\phantom{-}0.0000$ & 1.752 $\pm$ 0.007 & 1.774 $\pm$ 0.008 & 1.761 $\pm$ 0.005 & 1.756 $\pm$ 0.003 \\
\end{tabular}
\end{table}

\begin{table}[h]
\centering
\caption{\label{tab3} Fractal dimension for space--time quantum fractals defined in terms of the flux--driven (Bohmian) trajectories considered in this work.
These trajectories are associated with the wave function whose quantum carpet is shown in Fig.~1(a); their initial positions $x_0$ are indicated in the first column.}
\begin{tabular}{ccccc}
$x_0$ & haar & db4 & db8 & db16 \\
\hline
$-0.3750$ & 1.216 $\pm$ 0.010 & 1.167 $\pm$ 0.049 & 1.413 $\pm$ 0.284 & 1.168 $\pm$ 0.013 \\
$-0.3542$ & 1.255 $\pm$ 0.003 & 1.265 $\pm$ 0.028 & 1.248 $\pm$ 0.032 & 1.260 $\pm$ 0.014 \\
$-0.3333$ & 1.250 $\pm$ 0.001 & 1.247 $\pm$ 0.001 & 1.244 $\pm$ 0.001 & 1.250 $\pm$ 0.000 \\
$-0.3125$ & 1.245 $\pm$ 0.001 & 1.261 $\pm$ 0.014 & 1.247 $\pm$ 0.020 & 1.270 $\pm$ 0.002 \\
$-0.2917$ & 1.240 $\pm$ 0.005 & 1.248 $\pm$ 0.002 & 1.247 $\pm$ 0.003 & 1.247 $\pm$ 0.002 \\
$-0.2708$ & 1.254 $\pm$ 0.000 & 1.251 $\pm$ 0.004 & 1.248 $\pm$ 0.001 & 1.244 $\pm$ 0.004 \\
$-0.2500$ & 1.224 $\pm$ 0.015 & 1.228 $\pm$ 0.009 & 1.244 $\pm$ 0.005 & 1.242 $\pm$ 0.008 \\
$-0.2292$ & 1.256 $\pm$ 0.006 & 1.223 $\pm$ 0.009 & 1.223 $\pm$ 0.009 & 1.234 $\pm$ 0.010 \\
$-0.2083$ & 1.354 $\pm$ 0.166 & 1.323 $\pm$ 0.135 & 1.308 $\pm$ 0.089 & 1.446 $\pm$ 0.004 \\
$-0.1875$ & 1.245 $\pm$ 0.003 & 1.231 $\pm$ 0.010 & 1.269 $\pm$ 0.014 & 1.255 $\pm$ 0.000 \\
$-0.1667$ & 1.247 $\pm$ 0.003 & 1.253 $\pm$ 0.001 & 1.250 $\pm$ 0.001 & 1.266 $\pm$ 0.001 \\
$-0.1458$ & 1.235 $\pm$ 0.001 & 1.242 $\pm$ 0.004 & 1.238 $\pm$ 0.007 & 1.242 $\pm$ 0.006 \\
$-0.1250$ & 1.160 $\pm$ 0.026 & 1.052 $\pm$ 0.360 & 0.842 $\pm$ 0.384 & 1.258 $\pm$ 0.024 \\
\end{tabular}
\end{table}

\clearpage

\begin{table}[h]
\centering
\caption{\label{tab4} Fractal dimension for space quantum fractals profiles extracted from Fig.~1(b) at the times $t$ indicated in the first column.}
\begin{tabular}{ccccc}
time & haar & db4 & db8 & db16 \\
\hline
$T/\sqrt{2}$ & 1.533 $\pm$ 0.000 & 1.522 $\pm$ 0.007 & 1.498 $\pm$ 0.018 & 1.487 $\pm$ 0.018 \\
$T/\sqrt{3}$ & 1.461 $\pm$ 0.003 & 1.513 $\pm$ 0.000 & 1.500 $\pm$ 0.017 & 1.522 $\pm$ 0.000 \\
$T/\sqrt{5}$ & 1.544 $\pm$ 0.000 & 1.491 $\pm$ 0.000 & 1.545 $\pm$ 0.009 & 1.464 $\pm$ 0.000 \\
$T/\sqrt{7}$ & 1.439 $\pm$ 0.018 & 1.496 $\pm$ 0.010 & 1.520 $\pm$ 0.011 & 1.463 $\pm$ 0.026 \\
$T/\sqrt{10}$ & 1.513 $\pm$ 0.010 & 1.550 $\pm$ 0.048 & 1.518 $\pm$ 0.034 & 1.485 $\pm$ 0.016 \\
\end{tabular}
\end{table}

\begin{table}[h]
\centering
\caption{\label{tab5} Fractal dimension for time quantum fractals profiles extracted from Fig.~1(b) at the fixed $x-$positions indicated in the first column.}
\begin{tabular}{ccccc}
$x-$position & haar & db4 & db8 & db16 \\
\hline
$-0.4167$ & 1.804 $\pm$ 0.003 & 1.769 $\pm$ 0.009 & 1.751 $\pm$ 0.005 & 1.767 $\pm$ 0.012 \\
$-0.3333$ & 1.796 $\pm$ 0.010 & 1.770 $\pm$ 0.004 & 1.792 $\pm$ 0.026 & 1.775 $\pm$ 0.018 \\
$-0.2500$ & 1.793 $\pm$ 0.000 & 1.772 $\pm$ 0.009 & 1.771 $\pm$ 0.012 & 1.757 $\pm$ 0.007 \\
$-0.1667$ & 1.775 $\pm$ 0.002 & 1.752 $\pm$ 0.003 & 1.759 $\pm$ 0.010 & 1.784 $\pm$ 0.013 \\
$-0.0833$ & 1.781 $\pm$ 0.003 & 1.747 $\pm$ 0.004 & 1.798 $\pm$ 0.008 & 1.808 $\pm$ 0.015 \\
$\phantom{-}0.0000$ & 1.791 $\pm$ 0.000 & 1.777 $\pm$ 0.004 & 1.779 $\pm$ 0.005 & 1.777 $\pm$ 0.007 \\
\end{tabular}
\end{table}

\begin{table}[h]
\centering
\caption{\label{tab6} Fractal dimension for space--time quantum fractals defined in terms of the flux--driven (Bohmian) trajectories considered in this work.
These trajectories are associated with the wave function whose quantum carpet is shown in Fig.~1(b); their initial positions $x_0$ are indicated in the first column.}
\begin{tabular}{ccccc}
$x_0$ & haar & db4 & db8 & db16 \\
\hline
$-0.3750$ & 1.306 $\pm$ 0.042 & 1.266 $\pm$ 0.012 & 1.282 $\pm$ 0.016 & 1.314 $\pm$ 0.038 \\
$-0.3542$ & 1.278 $\pm$ 0.008 & 1.266 $\pm$ 0.019 & 1.273 $\pm$ 0.018 & 1.273 $\pm$ 0.016 \\
$-0.3333$ & 1.243 $\pm$ 0.025 & 1.265 $\pm$ 0.025 & 1.269 $\pm$ 0.025 & 1.280 $\pm$ 0.012 \\
$-0.3125$ & 1.263 $\pm$ 0.007 & 1.274 $\pm$ 0.020 & 1.268 $\pm$ 0.009 & 1.274 $\pm$ 0.017 \\
$-0.2917$ & 1.263 $\pm$ 0.011 & 1.274 $\pm$ 0.023 & 1.279 $\pm$ 0.020 & 1.283 $\pm$ 0.009 \\
$-0.2708$ & 1.263 $\pm$ 0.008 & 1.257 $\pm$ 0.023 & 1.257 $\pm$ 0.009 & 1.259 $\pm$ 0.004 \\
$-0.2500$ & 1.276 $\pm$ 0.022 & 1.314 $\pm$ 0.065 & 1.258 $\pm$ 0.031 & 1.264 $\pm$ 0.059 \\
$-0.2292$ & 1.261 $\pm$ 0.004 & 1.258 $\pm$ 0.013 & 1.258 $\pm$ 0.011 & 1.271 $\pm$ 0.014 \\
$-0.2083$ & 1.268 $\pm$ 0.010 & 1.258 $\pm$ 0.005 & 1.268 $\pm$ 0.007 & 1.270 $\pm$ 0.005 \\
$-0.1875$ & 1.249 $\pm$ 0.009 & 1.274 $\pm$ 0.022 & 1.270 $\pm$ 0.016 & 1.273 $\pm$ 0.018 \\
$-0.1667$ & 1.254 $\pm$ 0.009 & 1.267 $\pm$ 0.025 & 1.246 $\pm$ 0.051 & 1.275 $\pm$ 0.022 \\
$-0.1458$ & 1.275 $\pm$ 0.009 & 1.280 $\pm$ 0.011 & 1.266 $\pm$ 0.017 & 1.272 $\pm$ 0.016 \\
$-0.1250$ & 1.326 $\pm$ 0.178 & 1.304 $\pm$ 0.049 & 1.246 $\pm$ 0.095 & 1.253 $\pm$ 0.122 \\
\end{tabular}
\end{table}

\clearpage

\begin{table}[h]
\centering
\caption{\label{tab7} Fractal dimension for space quantum fractals profiles extracted from a quantum carpet where the initial wave function is a square covering the full well extension ($w = L$), as it was formerly considered by Berry \cite{berry:JPA:1996}, at the times $t$ indicated in the first column.}
\begin{tabular}{ccccc}
Time & haar & db4 & db8 & db16 \\
\hline
$T/\sqrt{2}$ & 1.559 $\pm$ 0.000 & 1.447 $\pm$ 0.023 & 1.455 $\pm$ 0.026 & 1.433 $\pm$ 0.000 \\
$T/\sqrt{3}$ & 1.407 $\pm$ 0.010 & 1.468 $\pm$ 0.010 & 1.399 $\pm$ 0.000 & 1.443 $\pm$ 0.000 \\
$T/\sqrt{5}$ & 1.504 $\pm$ 0.005 & 1.531 $\pm$ 0.006 & 1.558 $\pm$ 0.000 & 1.584 $\pm$ 0.000 \\
$T/\sqrt{7}$ & 1.386 $\pm$ 0.025 & 1.480 $\pm$ 0.010 & 1.506 $\pm$ 0.034 & 1.468 $\pm$ 0.025 \\
$T/\sqrt{10}$ & 1.433 $\pm$ 0.012 & 1.417 $\pm$ 0.022 & 1.484 $\pm$ 0.008 & 1.482 $\pm$ 0.026 \\
\end{tabular}
\end{table}

\begin{table}[h]
\centering
\caption{\label{tab8} Fractal dimension for time quantum fractals profiles extracted from a quantum carpet where the initial wave function is a square covering the full well extension ($w = L$), as it was formerly considered by Berry \cite{berry:JPA:1996}, at the $x-$positions indicated in the first column.}
\begin{tabular}{ccccc}
$x-$position & haar & db4 & db8 & db16 \\
\hline
$-0.4167$ & 1.749 $\pm$ 0.008 & 1.701 $\pm$ 0.010 & 1.722 $\pm$ 0.006 & 1.755 $\pm$ 0.012 \\
$-0.3333$ & 1.752 $\pm$ 0.010 & 1.775 $\pm$ 0.003 & 1.740 $\pm$ 0.007 & 1.734 $\pm$ 0.004 \\
$-0.2500$ & 1.762 $\pm$ 0.007 & 1.763 $\pm$ 0.003 & 1.748 $\pm$ 0.001 & 1.746 $\pm$ 0.008 \\
$-0.1667$ & 1.752 $\pm$ 0.001 & 1.749 $\pm$ 0.010 & 1.752 $\pm$ 0.010 & 1.752 $\pm$ 0.004 \\
$-0.0833$ & 1.741 $\pm$ 0.003 & 1.765 $\pm$ 0.011 & 1.765 $\pm$ 0.011 & 1.778 $\pm$ 0.005 \\
$\phantom{-}0.0000$ & 1.742 $\pm$ 0.008 & 1.742 $\pm$ 0.006 & 1.763 $\pm$ 0.012 & 1.752 $\pm$ 0.005 \\
\end{tabular}
\end{table}

\begin{table}[!h]
\centering
\caption{\label{tab9} Fractal dimension for space--time quantum fractals defined in terms of the flux--driven (Bohmian) trajectories considered in this work.
These trajectories are associated with the quantum carpet generated by an initial wave function described by a square function covering the full well extension ($w = L$), as it was formerly considered by Berry \cite{berry:JPA:1996}; their initial positions $x_0$ are indicated in the first column.}
\begin{tabular}{ccccc}
$x_0$ & haar & db4 & db8 & db16 \\
\hline
$-0.4167$ & 1.276 $\pm$ 0.000 & 1.255 $\pm$ 0.009 & 1.243 $\pm$ 0.006 & 1.332 $\pm$ 0.054 \\
$-0.3333$ & 1.244 $\pm$ 0.000 & 1.241 $\pm$ 0.010 & 1.243 $\pm$ 0.001 & 1.239 $\pm$ 0.003 \\
$-0.2500$ & 1.253 $\pm$ 0.007 & 1.252 $\pm$ 0.001 & 1.251 $\pm$ 0.002 & 1.244 $\pm$ 0.005 \\
$-0.1667$ & 1.245 $\pm$ 0.004 & 1.261 $\pm$ 0.009 & 1.263 $\pm$ 0.011 & 1.242 $\pm$ 0.007 \\
$-0.0833$ & 1.256 $\pm$ 0.001 & 1.248 $\pm$ 0.008 & 1.243 $\pm$ 0.006 & 1.257 $\pm$ 0.029 \\
\end{tabular}
\end{table}

\end{widetext}

\end{document}